# Critical dynamics in the evolution of stochastic strategies for the iterated Prisoner's Dilemma

Dimitris Iliopoulos[1#], Arend Hintze[1#] & Christoph Adami[1*]

[1]*Keck Graduate Institute for Applied Life Sciences, Claremont, CA 91711*

*email: adami@kgi.edu

[#]*These authors contributed equally to this work*

## Abstract

The observed cooperation on the level of genes, cells, tissues, and individuals has been the object of intense study by evolutionary biologists, mainly because cooperation often flourishes in biological systems in apparent contradiction to the selfish goal of survival inherent in Darwinian evolution. In order to resolve this paradox, evolutionary game theory has focused on the Prisoner's Dilemma (PD), which incorporates the essence of this conflict.

Here, we encode strategies for the iterated Prisoner's Dilemma (IPD) in terms of conditional probabilities that represent the response of decision pathways given previous plays. We find that if these stochastic strategies are encoded as genes that undergo Darwinian evolution, the environmental conditions that the strategies are adapting to determine the fixed point of the evolutionary trajectory, which could be either cooperation or defection. A transition between cooperative and defective attractors occurs as a function of different parameters such a mutation rate, replacement rate, and memory, all of which affect a player's ability to predict an opponent's behavior.

These results imply that in populations of players that can use previous decisions to plan future ones, cooperation depends critically on whether the players can rely on facing the same strategies that they have adapted to. Defection, on the other hand, is the optimal adaptive response in environments that change so quickly that the information gathered from previous plays cannot usefully be integrated for a response.

## Author Summary

The observed cooperation between genes, cells, tissues, and higher organisms represents a paradox for Darwinian evolution, because the individual success of cheating is rewarded before its long-term detrimental consequences are felt. The tension between cooperation and defection can be represented by a simple game (the "Prisoner's Dilemma"), which has been used to study the conflicts between decisions to cooperate or defect. Here, we encode these decisions within genes, and allow them to adapt to environments that differ in how well a player can predict how an opponent is going to play. We find that evolutionary paths end at strategies that cooperate if the environment is sufficiently predictable, while they end in defection in uncertain and inconsistent worlds because inconsistency favors defection over cooperation. This work shows that cooperation or defection, in populations of players that use the information from previous moves to plan future ones, can be influenced by changing the environmental parameters.

## Introduction

The evolution of cooperation is difficult to understand within Darwinian theory [1-3]. Indeed, cooperation is intrinsically vulnerable to exploitation because evolution rewards individual success, while any detrimental long-term effects for the group are secondary [4,5]. The tension between the short-term benefits of defection and the long-term benefits of cooperation has been studied using the Prisoner's Dilemma as a paradigm of social conflicts [3,6-9]. Previous work has shown that cooperation can only emerge in the presence of different enabling mechanisms. The main ones are direct reciprocity [6,10] (which can emerge when players play against each other repeatedly), spatial reciprocity [7], which is ensured if players only play neighbors on a regular grid (or more generally, on arbitrary graphs, giving rise to "network reciprocity" [11]), tag-based selection [12] (where players can recognize each other using some observable trait), kin selection [13], indirect reciprocity [14,15] (where cooperative or altruistic acts increase a player's reputation), or group selection [16]. Social diversity, where either the payoffs or the neighborhoods vary from player to player [17,18] can also enhance cooperation, as can "active linking" [19,20], where players differ in the rate at which they maintain interactions with other players. Generally speaking, the co-evolution of strategies with the different enabling mechanisms can also increase cooperation [21]. In all the discussed scenarios, a player's strategy is such that they either cooperate or defect in a deterministic manner, sometimes conditionally on previous plays.

If a cooperating strategy *accidentally* defects (or a defector accidentally cooperates) the noise that is introduced in this manner can have a dramatic effect on the competition. For example, among the (deterministic) strategies that take one previous move into account in order to decide how to play, the reciprocating strategy "TFT" (Tit-for-Tat) dominates [6], but is outcompeted [22-24] by "Win-Stay-Lose-Shift" (WSLS), which can correct for occasional mistakes [23]. Experiments with bacteria [25] and social amoeba [26] indeed suggest that the decision to cooperate or defect (in a general sense) is stochastic, and moreover that these decisions are controlled by genetically-encoded probabilities that are evolvable [27]. Rather than assuming that noisy decisions are either due to fuzziness in perception or lack of control over one's action [22], here we allow these probabilities to be fine-tuned by adaptation in response to the environment. We find that if a player's stochastic decisions are under genetic control, then the level of uncertainty about an opponent's next move (given their previous encounter) determines whether cooperation or defection evolves. Because this uncertainty is a direct consequence of environmental conditions, we conclude that when decisions are based on previous interactions, these conditions alone are sufficient to explain the evolution of cooperation in populations. Note that the stochasticity introduced by probabilistic play controlled by genes is fundamentally different from other random effects that can be introduced into evolutionary game dynamics, such as a probability to inherit a neighbor's strategy [28], or stochastically fluctuating payoffs [29,30], because neither of them can evolve.

In its simplest form, PD players have only two play options: cooperate (C) or defect (D). Both players are awarded a payoff R for mutual cooperation and a payoff P for mutual defection. Unequal moves award S to the cooperator and T to the defector. In standard PD [6], the values of the payoffs are constrained so that $T > R > P > S$ and $R > (S+T)/2$. The first equation ensures that for a single round of play, defection is an evolutionary stable strategy [4], while the second equation ensures that reciprocation of cooperation is favored over the trading of cooperative with defective moves. In the repeated PD (iterated PD or IPD) that we study here, two players meet more than once, and can establish cooperation by means of direct reciprocity [6]. In particular, we study exclusively the IPD with *memory*, that is, where players base their decision on previous plays (except for the first move with a new opponent). The term "memory" is not meant to imply that only higher organisms can engage in such strategies. Rather, stochastic decisions can be based entirely on the levels of protein on a cell's receptor, for example, and where these protein levels are the result of a cellular "decision" at an earlier time. A simple example for such a stochastic decision in response to the decision of other cells is quorum sensing in bacteria (see, e.g., [31]). We contend that the introduction of information exchange between players (via conditional strategies) is crucial for the evolution of cooperation.

## Results

We evolve strategies in spatially-structured and well-mixed finite populations, as it is known that the evolutionary dynamics depend on population structure as well as size (small fitness differences are effectively neutral only in finite populations [32]). Evolution experiments are carried out with populations on a regular 32x32 grid with wrapping boundary conditions, where the manner of replacement determines the population structure. Players engage their eight closest neighbors exactly once every update (playing one move), for 500,000 iterations. At the end of each update, a proportion $r$ (the replacement rate) of players is randomly eliminated using a Moran-like process [33,34], establishing a finite probability of future encounters between players beyond the first [35]. For spatially-structured populations, each player marked for death is replaced by an offspring of one of his neighbors, while for well-mixed populations the entire grid of players is considered for filling the empty position. In both population types, replicating players are chosen in proportion to their fitness, defined as the accumulated score. Scores are awarded according to the standard payoff matrix of Axelrod [35] throughout, with T=5, R=3, P=1, S=0.

For memory-one strategies, each player is represented by a genotype (strategy) composed of five genes, four of which encode the conditional probabilities $P_{XY}$ representing the probability that a player will cooperate, given that his last historical play was X and his opponent's response was Y, along with the unconditional probability $P_C$ to cooperate on the first move [24]. Each population is seeded with the "random" genotype where each of the five probabilities is set to 0.5. At each replication event, genes are subject to a per-gene mutation rate $\mu$, replacing that gene's probability to cooperate with a uniformly distributed number between 0 and 1.

For each evolutionary run, we record the genotype as well as phenotype (play statistics $\pi_{CC}$, $\pi_{CD}$, $\pi_{DC}$, and $\pi_{DD}$, given by the fraction of that type of play among all plays) for each organism on the line of descent (LOD) [36]. The LOD is generated by randomly selecting a genotype at the end of each run and tracing back its ancestry to the seeding genotype. Compared to the previously discovered deterministic memory-one strategies [37], our genetic implementation leads to the evolution of novel and drastically different successful strategies, depending on mutation rate, replacement rate, and population structure. None of the 32 deterministic strategies ever appear on the LOD, but instead, strategies evolve that are either cooperative or defective, depending on the experimental setting. Using the LOD averaged over 80 runs (see Fig. S1), we can obtain a consensus genotype for the particular experiment by averaging all genotypes in the latter half of this average LOD, removing any influence from the starting conditions (see Methods).

The consensus genotype for spatially-structured populations at low mutation and replacement rates is that of a cooperative strategy ($P_C$, $P_{CC}$, $P_{CD}$, $P_{DC}$, $P_{DD}$=0.647, 0.989, 0.234, 0.318, 0.448), as is evident from a commitment to exchange C plays (i.e., $P_{CC} \approx 1$) and a tendency to cooperate on the first move. By having a low $P_{CD}$ probability this strategy maintains a low tolerance to opponent defection and displays an unwillingness to be exploited. Maintaining a $P_{DD}$ value close to 0.5 with a slight bias towards defection, the consensus genotype expresses indifference in propagating defection but willingness to return to cooperation, a behavior not previously seen among stochastic strategies [24]. When faced with defective play, the strategy will acquire a deficit in lifetime payoff, which can be offset by exploiting naïve cooperators (and occasionally similar strategies) as indicated by a low $P_{DC}$ probability. Consensus strategies for cooperation in well-mixed populations, as well as defectors in both population structures (that appear at high mutation and replacement rates) are listed in Table S1 and described in Text S1.

In order to monitor the evolution of strategies, we reduce strategy space by performing a principal component analysis (PCA) of the probabilities on the average LOD obtained from 80 runs at mutation rate $\mu$=0.5% and replacement rate $r$=1%, and use these components to display the average trajectory at other mutation rates as well. For the spatially-structured population the first two principal components explain 83% and 10% of the variance, respectively (see Methods). Within the two-dimensional window defined by these principal components, we can also mark the location of some well-known strategies (see Fig. 1). We find that evolutionary trajectories obtained from the average LOD move towards a fixed point defined by a consensus genotype (see Methods) that represents the dominant strategy in the particular regime, while the actual genotypes on the LOD form a cloud in strategy space around the consensus that defines the strategy *attractor*. Strategies form clouds around this attractor because in a genetic implementation of IPD, the selective pressures acting on genes depend on the population a player finds himself in. For example, the DD gene in a cooperating population will begin to drift, only to return to its adaptive value when an invasion of defectors reinstates the selective pressure. Similarly, the CD and DC genes are under weakened selection in spatial populations because they are only expressed at the boundaries of homogeneous clusters.

The path in strategy space along the average LOD depends strongly on the mutation rate, and shows a qualitative switch—reminiscent of a phase transition—from the cooperative attractor RC (Fig. 1A) to the defecting attractor RD (Fig. 1C) at a critical value (Fig. 1B), as the mutation rate is increased. Studying the trajectories that emanate from the 16 (ignoring the first gene) deterministic strategies (Fig. 1D) suggests that the evolutionary fixed points are unique attractors for a given environment. We characterize the attractors with an *order parameter m* generated from the average play frequencies:

$$m = \frac{\langle \pi_{CC} \rangle - \langle \pi_{DD} \rangle}{\langle \pi_{CC} \rangle + \langle \pi_{DD} \rangle}, \quad\quad [1]$$

which is the normalized difference between frequencies of cooperative and defective play, averaged over the genotypes on the LOD after equilibration (see Methods). This parameter crosses zero at a critical mutation rate (Fig. 2A), indicating a transition from cooperative to defective strategies. We find that a transition can also be forced by changing the replacement rate *r*, as well as other parameters discussed below.

We can study the evolution of cooperation by plotting the order parameter Eq. [1] as a function of *r* and *µ* in a *phase diagram* that shows that both low replacement rate and low mutation rate lead to cooperation (Fig. 2B), but that the cooperative phase is much smaller for well-mixed populations (Fig. 2C). As *µ* approaches 0.5 (a per-genome mutation rate of 2.5 mutations per replication event), both the spatially-structured and the well-mixed populations begin to drift randomly, signaling that selection has become incapable of maintaining the genetic information. This transition is likely a *quasispecies delocalization* [38], but is smooth rather than abrupt owing to the small genome size [39]. That all strategies occur with equal frequencies in the population when taking the limit of very high mutation rate has been noted before [40].

Previous studies have only investigated small slices of this phase diagram by varying the average number of rounds between players [6] (for deterministic strategies) or varying the mutation rate in analytic calculations and numeric simulations of an infinitely iterated Markov process [11,40], concluding that cooperation is favored in spatially-structured population but not in well-mixed ones [41]. The phase diagram suggests instead that both cooperation and defection are possible in either population structure, but that the parameter range that facilitates cooperation in well-mixed populations is more restricted.

As the order parameter Eq. [1] is obtained from play statistics that represent the phenotype of players, we may ask how this transition is reflected in the genotype instead. The consensus genotype shows a marked decrease of the $P_{CC}$ probability as mutation rate increases, with clear differences between strategies in spatial (Fig. 3A) versus well-mixed (Fig. 3B) scenarios, as has been noted before [24]. At the critical mutation rate (Fig. 3, dashed vertical lines), the probability to cooperate after CC equals the probability to defect after DD. Thus, the consensus genotypes mirror the play statistics obtained to define the critical point.

## *Discussion*

Cooperation is inherently more risky than defection because it forgoes a guaranteed return (P) with the expectation of a benefit (R), rather than keeping the guaranteed return hoping for a windfall (T). This risk is mitigated if the uncertainty about receiving the benefit is reduced. For example, spatial reciprocity allows kin strategies to preferentially play each other (because kin place offspring close to themselves) thus increasing trust. In our model, an increase in mutation rate decreases the probability that kin play the same strategy (because mutations change the strategy of kin), and thus increases the uncertainty about the identity of the strategy a player will face. An increased replacement rate has a similar effect, as increasing $r$ shortens the average number of plays that a pair engages in, and this again decreases the probability to face a kin strategy (mutated or not). Previously, a general theory for the evolution of cooperation has been proposed [42,43] that posits that positive assortment between a player's genotype and the opponent's phenotype is sufficient to promote cooperation, using arguments that ultimately recapitulate Queller's [44] extension of Hamilton's rule.

In our experiments with stochastic conditional strategies, the assortment between a player's genotype and an opponent's phenotype is generated via the evolution of conditional interactions between the players [45], i.e., their ability to base their decisions on information about past behavior. In a sense, evolutionary adaptation creates this assortment by forging a "model" of the environment (in terms of the probabilities $P_{XY}$) that is adapted to the phenotype given by the play frequencies $\pi_{XY}$. For example, the cooperative fixed point represents a strategy that cooperates with cooperators, retaliates against defectors, but also forgives mistakes. Thus, it models an environment where cooperators dominate, errors happen, and sometimes defectors try to invade.

More uncertain environments reduce the accuracy of the model, thereby reducing positive assortment, leading to reduced cooperation. Can changing environmental conditions then drive a population from a cooperating to a defecting phenotype and vice versa? In Fig. 4, we show the order parameter of an adapting population on the line of descent where we changed the mutation rate abruptly from one favoring defection to one favoring cooperation, and back. We see that the population responds quickly (in terms of evolutionary time) and predictably to the changes.

If consistent environments enable cooperative behavior of strategies that rely on "sensing" their environment, we should also be able to influence the critical mutation rate (where cooperation turns into defection) by changing other parameters that affect uncertainty. For example, it is possible to increase player memory so that the last *two* moves by both players are taken into account to make decisions about cooperation or defection. In this case, player strategies are encoded in 21 genes, which can be used to predict future moves. As expected, the critical mutation rate is pushed to higher genomic mutation rates $\mu L$ (where $L$ is the number of genes) for memory 2 (Fig. 5A), and even higher for memory 3 (data not shown). Another source of unpredictability is the maximal strategy uncertainty given by the Shannon entropy [46] of the genome. In the present implementation, the probabilities that affect player decisions are coarse-grained to a resolution of 32,768 different alleles for each probability, or 15 bits of entropy per gene. Decreasing this resolution decreases the uncertainty generated by mutations. Fig. 5B shows the dependence of the order parameter on mutation rate for coarse-grainings of strategy space down to 1 bit (the deterministic strategies). In this limit, the critical mutation rate (for 1% replacement) is pushed towards $\mu=10\%$, implying that higher mutation rates result in defective play even though cooperation is expected [7,47]. Thus, obtaining more information about the environment, for example by basing decisions on more than one past move, increases the amount of information that a player can use to model the environment, and therefore gives rise to a more close assortment between genotype and opponent phenotype, which increases cooperation.

A framework where evolutionary game theory is implemented via genes that are under mutation and selection could also be used to predict how manipulation of the environment will affect the evolutionary fixed point in other systems. For example, defection has been observed in a number of biological systems whose dynamics can be described by a PD payoff matrix [48,49]. It is tempting to imagine that these systems can be coaxed into cooperation if mutation rate or turnover rate can be manipulated (as is shown in Fig. 4).

Evolution can be viewed as a process in which organisms increase their fit to the world by acquiring information about their environment [36,50]. Via this process, genomes become correlated to their environment, that is, genotypes that are adapted to their niche covary with the niche's character. Clearly, such a covariance is greatly enhanced if organisms can sense their environment, and thus base their decisions appropriately on the context. Therefore, we can expect that the evolution of sensory circuits that inform decisions should ultimately lead to a sufficient amount of covariance so that cooperation is expected according to Queller's rule [44] (unless environments are so *inherently* uncertain that they must remain uninformative to any player). If this is indeed true, then it appears that cooperation does not need to be added as a "third fundamental principle of evolution beside mutation and natural selection" as was suggested before [9], because it is a consequence of evolution.

## *Methods*

**Population dynamics.** The payoff for different moves was kept constant at Axelrod's values for all simulations:

$$\begin{pmatrix} R & S \\ T & P \end{pmatrix} = \begin{pmatrix} 3 & 0 \\ 5 & 1 \end{pmatrix}.$$

At each update, every player on the 32x32 grid (with wrapping boundary conditions) plays each of its neighbors exactly once. Upon birth, each player begins by consulting its $P_C$ gene for each opponent, and one of the four conditional genes thereafter, depending on its own play and the opponent's response. Players are selected for removal randomly with a probability given by the replacement rate $r$, giving rise to overlapping generations (asynchronous updating) [51,52]. As long as the player and its opponent are not replaced, they continue to consult their conditional genes to make decisions, so the replacement rate determines the average length of play history between two players (if a player's partner is replaced, the partner is greeted by consulting the unconditional gene). For most replacement rates, the first gene is consulted so rarely that it drifts neutrally, with a mean around 0.5 and a variance of 1/12, as expected for a uniformly distributed random variable bounded by zero and one. As a consequence, we often do not show any statistics for this gene.

To implement well-mixed populations using our grid structure, we only changed the identity of the pool used for replacing individuals marked for death, thus keeping the rest of the dynamics consistent. For structured populations, the eight neighbors of the marked individual are candidates for replication, with a probability proportional to their fitness given by their lifetime accumulated score. For well-mixed populations, the pool is given by all 1,023 remaining strategies in the population (in a Moran process, it is not usual for the individual to be marked for death to be included in the candidates for replication), but each strategy still plays eight neighbors. The player to be replaced, on the other hand, is chosen randomly among all 1,024 players in the population, irrespective of population structure or fitness. After replication, a genotype is mutated with a probability $\mu$, which is the mean number of mutations per gene per individual, implemented as a Poisson process. For most of the results in this study, the gene's probabilities are coarse-grained to 15 bits, which means that the probabilities are chosen from among $2^{15}=32,768$ possible values, representing the number of possible alleles at that locus. This resolution affects the critical mutation rates as shown in Fig. 5B, but increasing the resolution past 15 bits does not (data not shown). Because the mutation probabilities are thought to represent the decision of entire pathways of perhaps hundreds of genes, they should not be compared to per-nucleotide mutation rates.

**Line of descent and consensus genotypes.** Rather than collecting population averages of plays, we instead study the evolution of strategies by following the line of descent (LOD) of player genotypes for each replicate run. The LOD is obtained by choosing a random player at the end of the run and following its direct ancestors backwards to the first genotype [36]. Fig. S2A shows a typical sequence of genotypes, while Fig. S2B shows the play statistics for the same LOD. The population average of play statistics for the same experiment is shown for comparison in figure S2C. Average lines of descent and average play statistics along the line of descent can be created by averaging, for each update, the probabilities of the genotypes as well as the probabilities of play, of the organism on the LOD of each of the 80 replicates at that update. Fig. S1A shows such an average genetic LOD, while figs. S1B and C show the average play statistic on the line of descent for two different mutation rates. The latter two figures show that the average play statistics converge towards evolutionary fixed points that we term the *consensus genotype*, but that the time to achieve this fixed point depends on the mutation rate. The consensus genotype for each set of replicates is obtained by averaging the second half of the average genetic LOD minus the last 50,000 updates, which removes most or all of the transient and also the variance due to picking random genotypes as the originators of the LOD. Indeed, because the LOD splits at the most recent common ancestor (MRCA) of the population at the end of the run, the LOD past the MRCA is not necessarily representative of the evolutionary dynamics (as seen for example in Fig. S1B.) Discarding the last 50,000 updates truncates the LOD to genotypes before the MRCA for almost all runs. Using the MRCA genotype instead of the consensus genotype as representative of the fixed point does not change the results.

**Principal Component Analysis.** We create the evolutionary trajectories in Fig. 1 and Fig. S3 by performing a principal component analysis of the set of probabilities ($P_{CC}$, $P_{CD}$, $P_{DC}$, $P_{DD}$) from all of the 500,000 data points on the average genetic LOD of the 80 replicates at mutation rate $\mu$=0.5% and replacement rate $r$=1%, for both the spatially-structured and the well-mixed population, respectively. Because the first gene ($P_C$) is consulted so rarely it drifts almost neutrally and is for that reason omitted from the PCA. Including it does not significantly affect the four other principal components (data not shown). For the spatially structured population we obtain $PC_1$=(-0.86, 0.192, -0.055, -0.47) and $PC_2$=(-0.348, 0.442, -0.065, 0.824). These components explain 83% and 10% of the variance respectively. For the well-mixed population, the principal components

are $PC_1$=(-0.714, 0.132, -0.162, -0.668) and $PC_2$=(-0.393, 0.54, 0.646, 0.37), explaining 86% and 7% of the variance, respectively. To depict the evolutionary trajectories at higher mutation rate (panels B and C in Fig. 1 and panels B-D in Fig. S3), we keep the principal components obtained with the low mutation rate strategies so that the landmarks given by the common deterministic strategies such as TFT (Tit-for-Tat), WSLS (Win-Stay-Lose-Shift), ALL-C, and ALL-D remain at the same positions. These fixed components are also used to plot the location of the consensus genotype at mutation rate 0.5% (RC, the "robust cooperator"), and the consensus genotype at mutation rate 5% (RD, the "robust defector"). The consensus strategies RC and RD for spatially-structured and well-mixed populations are different, and described in the supplementary text below. Using the principal components implied by the average LOD obtained at 5% mutation rate (defecting attractor) instead does not change the nature of the results (data not shown).

## Acknowledgements

We thank Bjørn Østman and Sharmila Kopanathi for discussions, as well as Richard Lenski and Claus Wilke for comments on the manuscript. This work was supported by the NSF' Frontiers in Integrative Biology Grant FIBR-0527023.

## Figures

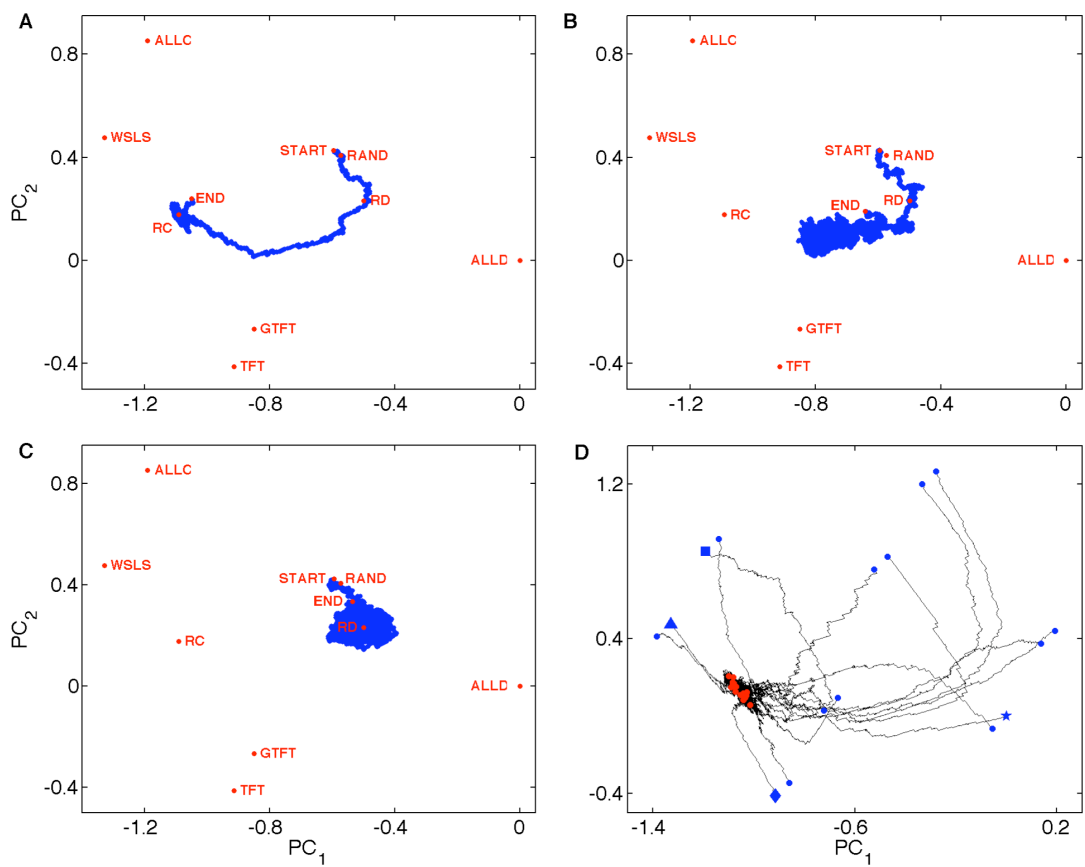

**Fig. 1. Evolutionary trajectories and attractors**. All trajectories start at the same point (START), and move towards the strategy marked by 'END'. Several well-known strategies provide landmarks in strategy space: TFT: $(P_{CC}, P_{CD}, P_{DC}, P_{DD})$=(1,0,1,0), ALLC=(1,1,1,1), ALLD=(0,0,0,0), WSLS=(1,0,0,1), GTFT=(1,0.333,1,0), START=(0.5,0.5,0.5,0.5). All experiments shown are run in a spatially-structured

environment at replacement rate $r$=1%. Trajectories for well-mixed populations are shown in Fig. S3. (**A**) Evolution of the average LOD for $\mu$=0.5%. RC marks the consensus genotype (see Methods) of this trajectory, while RAND marks the consensus genotype at $\mu$=50%, when the population drifts neutrally. This attractor is not the same as 'END' because that genotype lies past the most recent common ancestor of the population. (**B**) Trajectory for $\mu$=2.5%, close to the critical mutation rate. (**C**) Trajectory for $\mu$=5%. 'RD' marks the consensus genotype for these parameters. (**D**) Trajectories emanating from 16 deterministic strategies (at $\mu$=0.5%) suggest that the fixed point is unique. Blue symbols: start, red dots: end points. Symbols: ◆: TFT, ■: ALLC, ✶: ALLD, ▲: WSLS. **A-D** use principal components of trajectory shown in **A**.

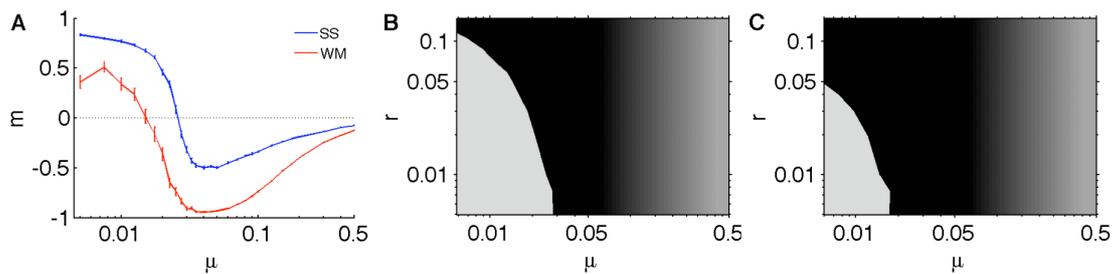

**Fig. 2. Transitions in strategy space**. (**A**) The order parameter $m$ defined in Eq. [1] as a function of the mutation rate for a spatially-structured and a well-mixed population, obtained from play statistics averaged over 80 independent runs each (see Text S2). Errors are two standard errors. (**B**) Qualitative phase diagram as a function of $\mu$ and $r$ for spatially-structured populations, where light grey indicates cooperation and black indicates defection. (**C**) Phase diagram for well-mixed populations. Both phase diagrams with quantitative levels of cooperation are shown in Fig. S4.

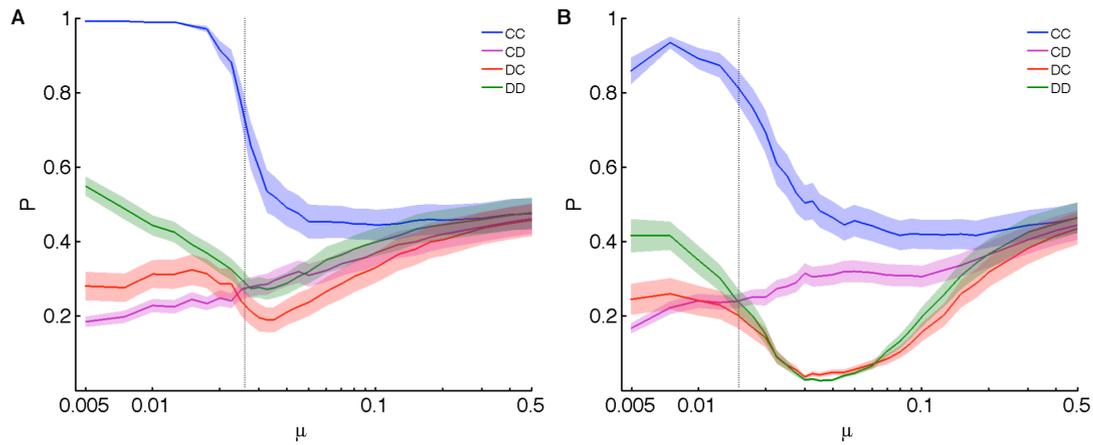

**Fig. 3. Evolution of consensus genotypes**. Mean of probabilities of the consensus genotype as a function of mutation rate ($r$=1%). Colored areas represent the variance of the probability distribution, and reflect the strength of selection. ($P_C$ is omitted because it drifts neutrally, see Methods). Vertical lines drawn at the critical mutation rate. (**A**) Spatially-structured. (**B**) Well-mixed.

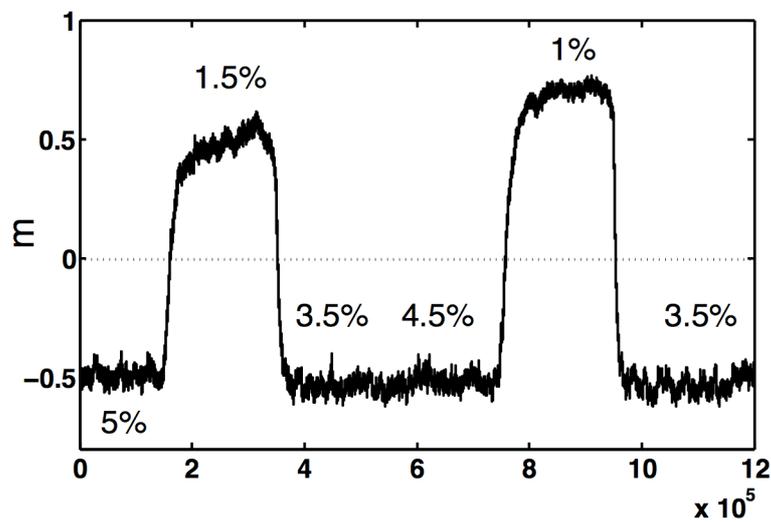

**Fig. 4. Strategy evolution under changing mutation rates.** Order parameter $m$ as a function of update time for an experiment with five changes in mutation rate, starting with a type adapted to a high mutation rate of 5% (defection regime). We show the order parameter for the average LOD of 80 runs with the same regime of mutation rate

changes. The population reacts to a changed mutation rate quickly, and settles around the fixed point appropriate for that mutation rate, indicated in the figure.

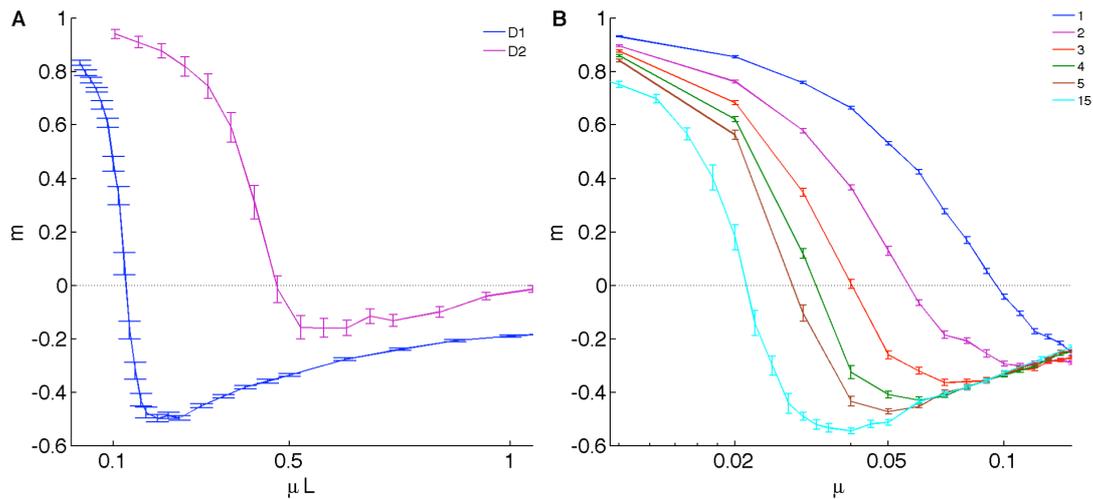

**Fig. 5. Order parameter in different environments for spatially-structured populations**. **(A)** Phase transition for populations playing with memories of different size as a function of genomic mutation rate *μL*, where *L*=5 for memory-one strategies (D1, blue line) and *L*=21 for memory-two strategies (D2, pink line). **(B)** Phase transition for environments with different resolutions of strategy space, from 15 bits per gene to 1 bit per gene (deterministic strategies). Colors as in legend.

## Supporting Information

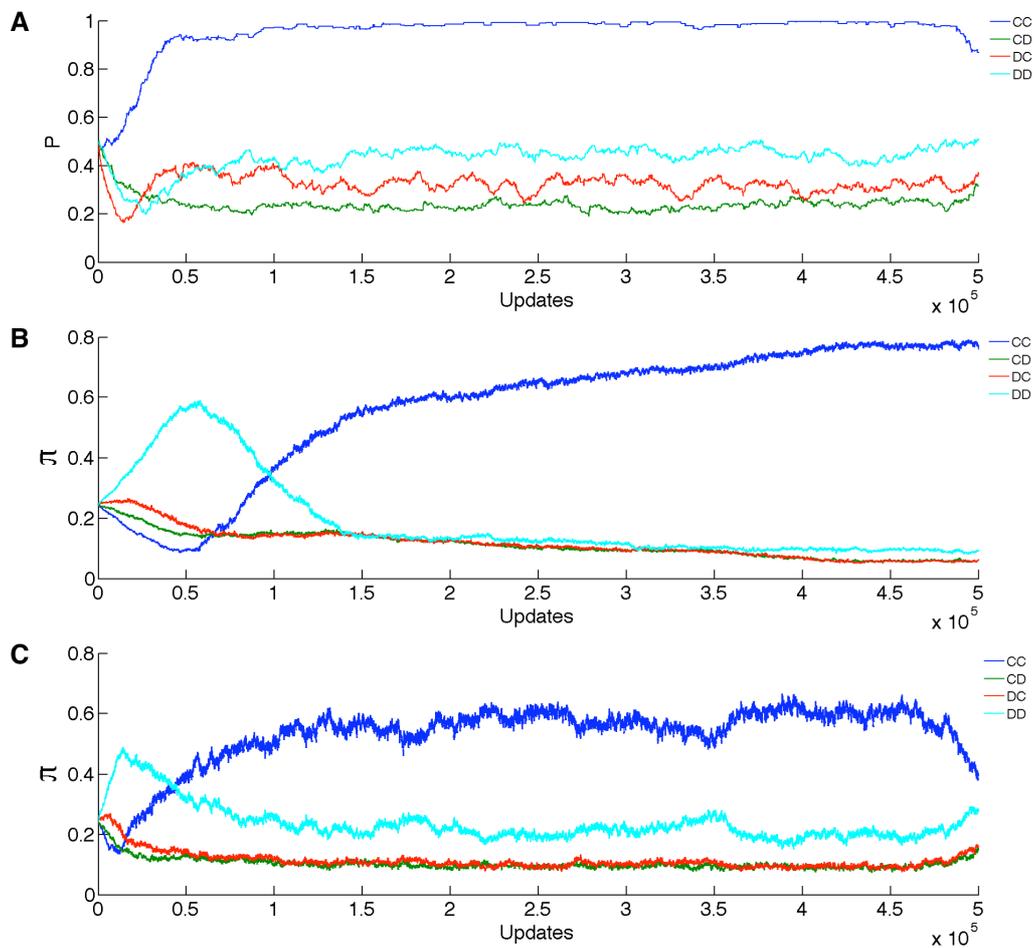

**Figure S1. Average gene probabilities and play statistics**. LOD gene probabilities ($P_{XY}$) and play statistics ($\pi_{XY}$) for a spatially-structured population, averaged over 80 experiments (500,000 updates each), at different $\mu$ and fixed $r$ (1%). $P_C$ and $\pi_C$ are omitted because $P_C$ drifts almost neutrally (see Methods) (**A**) Average gene probabilities recorded at $\mu$=1%. (**B**) Play statistics recorded at $\mu$=0.1%. (**C**) Play statistics recorded at $\mu$=2%.

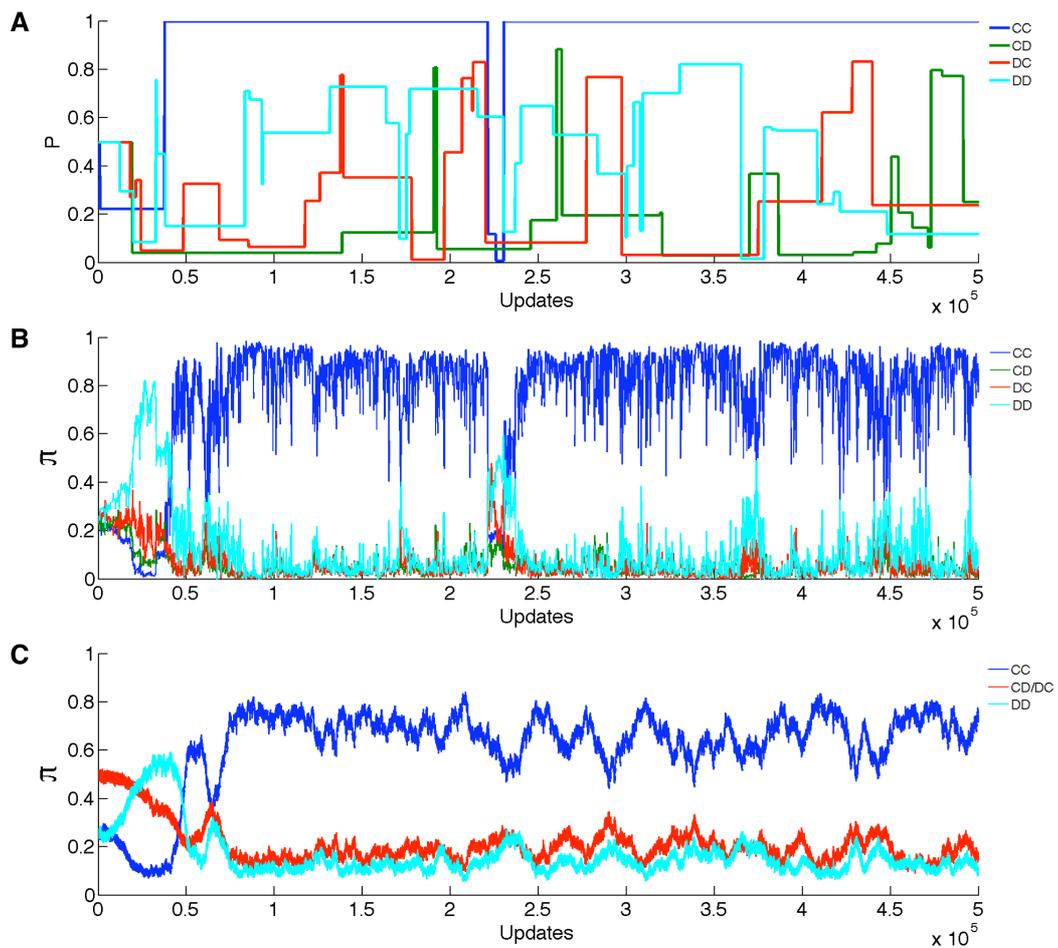

**Figure S2. LOD and population genotypes and phenotypes.** Single run LOD gene probabilities ($P_{XY}$) and play statistics ($\pi_{XY}$), as well as population average play statistics, for a spatially-structured population at $\mu$=1% and $r$=1%. $P_C$ and $\pi_C$ are omitted because $P_C$ drifts neutrally (see Methods). **(A)** LOD gene probabilities. **(B)** LOD play statistics. **(C)** Average population play statistics.

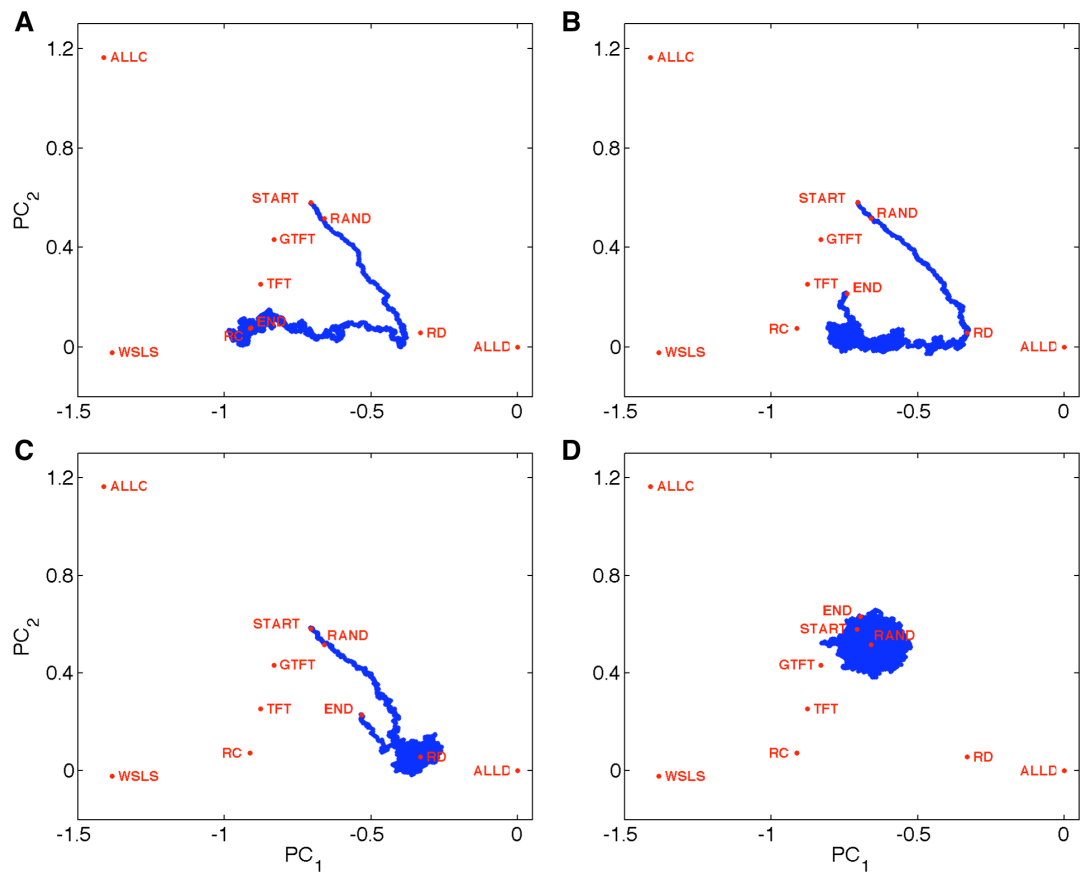

**Figure S3. Evolutionary trajectories and attractors for well-mixed populations.** All trajectories start at the same point ('START'), and move towards the strategy marked by 'END'. Several well-known strategies provide landmarks in strategy space: 'TFT': ($P_{CC}$, $P_{CD}$, $P_{DC}$, $P_{DD}$)=(1,0,1,0), 'ALLC'=(1,1,1,1), 'ALLD'=(0,0,0,0), WSLS=(1,0,0,1), GTFT=(1,0.333,1,0), START=(0.5,0.5,0.5). All experiments shown are run at replacement rate $r$=1% for well-mixed populations. (**A**), Evolution of the average LOD for $\mu$=0.5%. RC marks the consensus genotype of this trajectory (described in supplementary text). This attractor is not the same as 'END' because that genotype lies past the most recent common ancestor of the population. (**B**) Trajectory for $\mu$=1.5%, close to the critical mutation rate. (**C**) Trajectory for $\mu$=5%. 'RD' marks the consensus genotype for these parameters. (**D**) Trajectory for $\mu$=50%. 'RAND' marks the consensus genotype for these parameters.

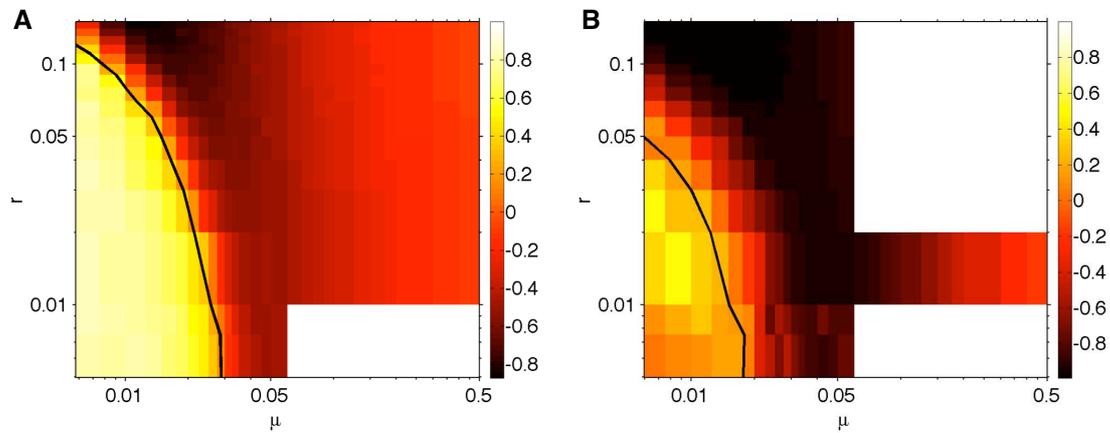

**Figure S4. Quantitative phase transition diagrams as a function of $\mu$ and $r$.** Coloring is applied according to the order parameter (*m*) with dark red to black indicating defection ($m < -0.2$), light yellow to light orange indicating cooperation ($m > 0.2$) and orange indicating a transition regime of equal cooperation and defection ($0.2 \geq m \geq -0.2$). White colored areas contain no recorded data. **(A)** Spatially-structured, **(B)** well-mixed.

**Text S1. Description of consensus strategies**

At low mutation rates (<1%) and a fixed replacement rate of 1%, robust cooperation quickly emerges as the dominant strategy for both well-mixed and spatially-structured populations. Robustness is generally regarded as a measure of a system's performance in the face of perturbation or uncertainty. In this context, robust cooperation describes three basic behaviors that allow players to play well against defective players and even when faced with their own strategy, described in the main text.

The consensus genotypes at high mutation rates (5%) and a fixed replacement rate of 1%, for both population types, are representative of robust defection. Robustness in this case is less of a mandate in a cooperative regime, since cooperation is inherently a more risky behavior in PD. Robust defectors have to maintain D exchanges while at the same time be willing to bait cooperative strategists with C plays and exploit them in the process. A $P_{CC}$ probability close to 0.5 with a bias towards defection compared to the $P_{CC}$ of RC is indicative of the unwillingness to engage in extended cooperative exchanges but at the same time a willingness to establish limited CC exchanges in hopes of future exploitation of a cooperative strategist. On the other hand, a very low $P_{DD}$ probability expresses eagerness to maintain defective play. For the well-mixed population, the $P_{DD}$ probability is much smaller than the equivalent probability in the spatially-structured population, which might be due to the absence of clusters under well-mixing giving rise to a higher degree of uncertainty experienced by players. The same reasoning might be applicable to the low $P_{DC}$ probability. Defector strategists in well-mixed populations are more willing to take advantage of cooperative players since they play against many more such players due to the absence of cooperative cluster shielding.

**Text S2. Experimental statistics**

For the spatially-structured phase transition experiments a total of 532 pairs of different mutation and replacement rate experiments were run, for a minimum of 13 and a maximum of 80 replicates for each pair (depending on the replacement rate), leading to 14,576 replicates in total. For the well-mixed phase transition experiments a total of 304 pairs of different mutation and replacement rate experiments were run, for a minimum of 10 and a maximum of 80 replicates each, leading to 10,080 replicates in total. For both population types, each experiment was run for 500,000 updates leading to approximately 7.3 billion updates for spatially-structured and 5 billion updates for well-mixed populations. All experiments were run in parallel on up to 25 dual core 2.6GHz (Intel® Pentium® Processor E5300) computers for an approximate total of about 8,500 hours for the spatially-structured and 5,900 hours for well-mixed population experiments. Data collected from the phase transition experiments were used to generate Figures 1, 2 and 3 as well as figures S1, S2, S3 and S4. For Figure 4A and memory depth 2, 16 different mutation rates were run at 1% replacement rate (80 replicates each), for a total of 1,280 experiments requiring 1,280 CPU hours. For Figure 4B, bits 1 to 5 were run for 40 replicate experiments at 15 different mutation rates (and a fixed replacement rate of 1%), for a total of 3,000 experiments (1,000 CPU hours). In total, the study required almost 29,000 individual experiments requiring a total of 1.9 CPU years (about 10 CPU weeks in parallel).

**Table S1. Consensus genotypes for different mutation rates and population structures.** Mean probabilities for each gene averaged over 80 average LODs, with variance in brackets. SS: spatially-structured population, WM: well-mixed population, COOP: cooperator, DEFEC: defector.

|  | $P_C$ | $P_{CC}$ | $P_{CD}$ | $P_{DC}$ | $P_{DD}$ |
|---|---|---|---|---|---|
| SS COOP ($\mu$ =1%) | 0.647 (0.088) | 0.989 (0.005) | 0.234 (0.035) | 0.318 (0.075) | 0.448 (0.054) |
| SS DEFEC ($\mu$ =5%) | 0.481 (0.084) | 0.458 (0.091) | 0.315 (0.062) | 0.243 (0.073) | 0.325 (0.064) |
| WM COOP ($\mu$ =1%) | 0.595 (0.098) | 0.893 (0.056) | 0.247 (0.038) | 0.247 (0.079) | 0.356 (0.076) |
| WM DEFEC ($\mu$ =5%) | 0.442 (0.081) | 0.460 (0.084) | 0.325 (0.059) | 0.063 (0.018) | 0.053 (0.012) |